\documentclass[12pt]{article}
\usepackage{latexsym}

\def\bs{\begin{subequations}}
\def\es{\end{subequations}}

\catcode`\@=11

\newtoks\@stequation
\def\subequations{\refstepcounter{equation}
  \edef\@savedequation{\the\c@equation}%
  \@stequation=\expandafter{\theequation}
  \edef\@savedtheequation{\the\@stequation}
  \edef\oldtheequation{\theequation}%
  \setcounter{equation}{0}%
  \def\theequation{\oldtheequation\alph{equation}}}

\def\endsubequations{\setcounter{equation}{\@savedequation}%
  \@stequation=\expandafter{\@savedtheequation}%
  \edef\theequation{\the\@stequation}\global\@ignoretrue}

\makeatletter
        \renewcommand{\theequation}{\thesection.\arabic{equation}}%
        \@addtoreset{equation}{section}%
\makeatother

\renewcommand{\thefootnote}{\fnsymbol{footnote}}

\begin{document}

\begin{titlepage}

 November 15, 2017 (version 2 of arxiv:1710.09904)

\begin{center}        \hfill   \\
            \hfill     \\
                                \hfill   \\

\vskip .25in

{\large \bf Tachyon Dynamics - for Neutrinos?\\}

\vskip 0.3in

Charles Schwartz\footnote{E-mail: schwartz@physics.berkeley.edu}

\vskip 0.15in

{\em Department of Physics,
     University of California\\
     Berkeley, California 94720}
        
\end{center}

\vskip .3in

\vfill

\begin{abstract}

Following earlier studies that provided a consistent theory of kinematics for tachyons (faster-than-light particles) we here embark on a study of tachyon dynamics, both in classical physics and in the quantum theory. Examining a general scattering process we come to recognize that the labels given to "in" and "out" states are not Lorentz invariant for tachyons; and this lets us find a sensible interpretation of negative energy states. For statistical mechanics, as well as for scattering problems, we study what should be the proper expression for density of states for tachyons. We review the previous work on quantization of a Dirac field for tachyons and go on to expand earlier considerations of neutrinos as tachyons in the context of cosmology. We stumble into the realization that tachyon neutrinos would contribute to gravitation with the opposite sign compared to tachyon antineutrinos. This leads to the gobsmacking prediction that the Cosmic Neutrino Background, if they are indeed tachyons, might explain both phenomena of Dark Matter and Dark Energy. This theoretical study also makes contact with the anticipated results from the experiments KATRIN and PTOLEMY, which focus on beta decay and neutrino absorption by Tritium.

\end{abstract}

\vfill

\end{titlepage}

\renewcommand{\thefootnote}{\arabic{footnote}}
\setcounter{footnote}{0}
\renewcommand{\thepage}{\arabic{page}}
\setcounter{page}{1}

\section{Introduction}

After a series of papers in which I developed consistent theoretical formulations for the kinematics of free tachyons (faster-than-light particles), \cite{CS1,CS2,CS3,CS4} here I will present some significant, if not final, work on the larger problem of tachyon dynamics.

Sections 2 and 3 deal with the prominent question of how to treat negative energy states of tachyons - first as classical point particles and then as quantum systems. This is done in the context of a general reaction/scattering process; and we learn that the familiar labels for "in" and "out" states have Lorentz invariant meaning only for slow particles, $v \le c$. For tachyons, we need only attend to a careful bookkeeping process to see that there is no problem with the physical laws nor with the standard precepts of Special Relativity.

In Section 4 the topic is density of states, as that concept is key in the study of statistical mechanics and also appears in scattering problems. Section 5 reviews recent work on quantizing a Dirac field theory of tachyons, which is thought to be potentially a description of neutrinos. In this we recognize the need to avoid "canonical" procedures that were developed for slow particles yet are able to arrive at a consistent resolution of the "problem" of causality, which is often invoked to reject the idea of tachyons. This contributes an alternative answer to the previous discussion of density of states; and in Section 6 we delve deeper into statistical mechanics for tachyons.

Section 7 goes into a review of current Cosmology theory with the inclusion of tachyonic neutrinos. This lets me improve the recent paper \cite{CS4}, which presented the surprising claim that if Cosmic Background Neutrinos are actually tachyons, then that can explain the phenomenon called Dark Energy. Additionally, the Friedmann Robertson Walker framework of cosmic evolution is expanded to include the possibility of tachyons in what appears to be a plausible alternative to the standard ($\Lambda$CDM) model.

In Section 8 I look more closely at the quantized field construction of the energy-momentum tensor for spinor tachyons and find a surprising result. The tachyonic neutrinos (of one helicity) and the antineutrinos (of the other helicity) contribute with opposite signs as sources of gravitation. Thus my two conflicting former hypotheses - that the tachyonic Cosmic Neutrino Background (CNB) could explain Dark Matter and that it could explain Dark Energy - may both be true.

In Section 9 I look at two imminent experiments, KATRIN and PTOLEMY, that are planned to look at the effects of massive neutrinos on the weak interaction of Tritium and make predictions about how those experimental results would look if the neutrinos are actually tachyons.

In Section 10 I summarize the new results found here and note some outstanding questions. Two appendices explore some issues that are not yet fully resolved.

\section{Classical Particles}
Let's look at a space-time diagram for a general interaction process. See Figure 1: time increases upward and space coordinates go out in the other three dimensions, of which we draw only one.
The black circle in the center is the (complicated) interaction region; the rest of the picture is about the asymptotic states, a collection of free particles going into or coming out of the interaction. The thick lines are the lightcones; the thin lines in Figure 1 show four particles involved in the reaction, 
$n \rightarrow p + e + \nu$, where I imagine that the neutrino is a tachyon.

\begin{picture}(100,200)
\thicklines 
\put(50,100){\circle*{10}}
\put(55,100){\line(1,1){60}}
\put(55,100){\line(1,-1){60}}
\put(45,100){\line(-1,1){60}}
\put(45,100){\line(-1,-1){60}}
\thinlines 
\put(50,105){\line(0,1){40}}
\put(50,95){\line(0,-1){40}}
\put(50,100){\line(-1,2){20}}
\put(55,100){\line(5,1){40}}
\put(50,150){\shortstack{p}}
\put(48,45){\shortstack{n}}
\put(25,145){\shortstack{e}}
\put(97,105){\shortstack{$\nu$}}
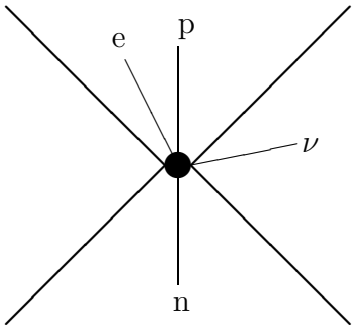
\put(0,20){\shortstack{Figure 1. Reaction with an outgoing tachyon.}}
\end{picture}

The upper cone contains the trajectories of all the outgoing ordinary particles and the lower cone contains the incoming ordinary particles.The asymptotic trajectories of any tachyons involved in this interaction will appear in the sides of this picture, outside of the lightcones, in areas that have a space-like distance from the interaction. In this picture, Figure 1,  we see a very low energy neutrino-tachyon ($E << m$) coming out. A higher energy tachyon would move close to the lightcone.

In terms of the coordinates $\xi^\mu(\tau)$, we look at the 4-vector velocity $\dot{\xi}^\mu (\tau) = (dt/d\tau, d\textbf{x}/d\tau)$ for any particle. For ordinary particles, we define the first component $dt/d\tau$ to be positive, and will identify this with the energy of the particle: $p^\mu = m \dot{\xi}^\mu = (E, \textbf{p})$.  Since these are, for ordinary particles, time-like 4-vectors that sign choice will be invariant under any proper Lorentz transformation. But for tachyons we must deal with the mathematical fact that $\dot{\xi}^\mu$ is a spacelike 4-vector and so the sign of its first component may change under a Lorentz transformation. 

Look at Figure 2. Is this a picture of the reaction $ n \rightarrow p + e + \nu$ with the neutrino carrying off negative energy; or is this a picture of the reaction $n + \nu \rightarrow p + e$ with positive energy for all participants?

\begin{picture}(100,200)
\thicklines 
\put(50,100){\circle*{10}}
\put(55,100){\line(1,1){60}}
\put(55,100){\line(1,-1){60}}
\put(45,100){\line(-1,1){60}}
\put(45,100){\line(-1,-1){60}}
\thinlines 
\put(50,105){\line(0,1){40}}
\put(50,95){\line(0,-1){40}}
\put(50,100){\line(-1,2){20}}
\put(55,100){\line(5,-1){40}}
\put(50,150){\shortstack{p}}
\put(48,45){\shortstack{n}}
\put(25,145){\shortstack{e}}
\put(97,90){\shortstack{$\nu$}}
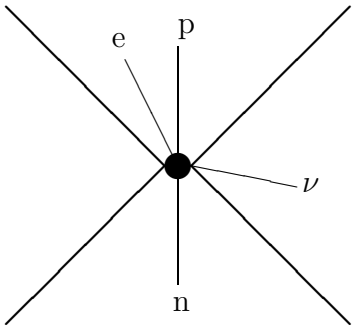
\put(0,20){\shortstack{Figure 2. Reaction with an incoming tachyon.}}
\end{picture}

This "problem" is the same as noting that the mass-shell equation,
\begin{equation}
p^\mu \; p_\mu = E^2 - p^2 = \pm m^2, \label{b1}
\end{equation}
gives us two separate hyperboloids for ordinary particles (plus sign) but a single hyperboloid for tachyons (minus sign). For ordinary particles, we manage to reinterpret the negative energy solutions as antiparticles and give them positive energy. For tachyons we need to see what it means when we look at a positive energy particle from a different Lorentz frame, where it appears to have negative energy.

Look again at the space-time diagrams above. A "positive energy" tachyon will have $dt/d\tau > 0$ and its trajectory will be seen moving upward - as in Figure 1 - and so we would call that an outward going particle if it sits above the interaction region in time; and we would call it an inward moving particle if the trajectory sits below the interaction region - as in Figure 2. But from another reference frame we may have $dt/d\tau < 0$ for what was formerly an outgoing particle and so this will now look like an incoming particle. The lesson is that \emph{the labels "in" and "out" are Lorentz invariant for ordinary particles but NOT for tachyons.}  Does this matter? No. The physical law which we call the conservation of total energy and momentum is written,
\begin{equation}
\sum_{out} p^\mu _j  - \sum_{in} p^\mu_i = 0.\label{b2}
\end{equation}
This is true in any Lorentz frame, even though the individual terms in this equation each transform. What happens to the momentum 4-vector of an individual tachyon when the Lorentz transformation changes the sign of its time-component? We simply move it from one group ("in" or "out") to the other: the energy always comes out positive, although the direction of the momentum will be reversed. This is consistent with how we read the trajectories in the figures above. Also, the helicity does not change.

Note. Some earlier authors \cite{ER}  writing about tachyons, have postulated this as a physical principal, called it a "reinterpretation", and even likened it to the  familiar reinterpretation of negative energy solutions of the Dirac equation.  But I emphasize that this is nothing more than a direct reading of what we say about the trajectories drawn in the space-time diagram for any interaction. We read the space-time diagram with time increasing upward. If we plot a trajectory on this diagram we say that it is incoming or outgoing depending on how we read that diagram.

\section{Quantum Theory}
We want to build up to an extension of S-Matrix theory so that it can encompass tachyons as well as ordinary particles. Let's start with the asymptotic states of all particles involved in some process.  We have to talk about two sets of data, which we arrange in two vectors belonging to a big Hilbert space.
I will write this as
\begin{equation}
\{\Psi_{out}; \Psi_{in}\}, \;\;\; \Psi_{in} = \prod_i\;|p_i^\mu>, \;\;\; \Psi_{out} = \prod_j\;|p_j^\mu>,\label{c1}
\end{equation}
where I use the standard notation $| p>$ for a ket vector representing a free particle of momentum $p$.  I use the notation $\{ ...; ...\}$ to denote the set of two vectors in the Hilbert space. This is not an inner product (that will come later).

We can talk about a Lorentz transformation and how it affects this set of two state vectors. For ordinary particles in each $\Psi$, the momentum vectors transform in the usual way. For any tachyons, the momentum 4-vector is transformed in the usual way, AND, if this produces a change in sign of the time component, then that whole one-particle ket is moved from its original $\Psi$ to the other $\Psi$, along with the change $p^\mu \rightarrow -p^\mu$.

This set of rules is simply good bookkeeping, conforming to what we have learned about the classical description of particles going into and coming out of a general interaction process.

Now we must get beyond the asymptotic states and start talking about the interaction.  For ordinary particles we have the whole machinery of Hamiltonian dynamics, first classical and then quantum mechanical, which leads up to the construction of the S-matrix,
\begin{equation}
S = e^{-i H t}, \;\;\;\;\; or\; S = T\;exp( -i \int_{-\infty}^\infty  H \;dt).\label{c2}
\end{equation}
While that is all nice for ordinary particles, which always propagate within the lightcone, this is of questionable value for tachyons. [See my discussion of this in a recent paper.\cite{CS3}] We need a more inclusive way of describing how the state vectors for a physical system will evolve as the constituent particles move and interact.

Let me assume that we have a Poincare invariant operator A (the action?) that gives the correct and complete propagation for the whole system of physical particles with their interactions in the form of an exponential 
\begin{equation}
\lim_{N \rightarrow \infty} (1+\frac{1}{N}\;\ iA)^N = e^{iA}. \label{c3}
\end{equation}

We then construct the transition probability amplitude
\begin{equation}
<\Psi^\dagger_{out}, e^{iA} \Psi_{in}>,\label{c4}
\end{equation}
where now we do mean an inner product, as usual, involving the Hermitian adjoint of the out state vector.

Also assume that we can write $A = A_0 + A_1$, where $A_0$ covers the free motion of all particles and $A_1$ has their interactions.  To first order we have the "perturbation series" expansion,
\begin{equation}
e^{iA_0 + iA_1} = e^{iA_0} + \int_0^1 ds e^{iA_0 (1-s)}\; iA_1\;e^{iA_0 s} + ...\label{c5}
\end{equation}

How does the free propagator $e^{iA_0}$ act upon the asymptotic state vector $\Psi_{in}$?
I think it just gives it a phase factor - like moving any plane wave function $e^{-ik_\mu x^\mu}$ through some distance in spacetime. That seems to give us a simple and familiar rule for first order transition amplitudes,
\begin{equation}
T = <\Psi_{out}, iA_1\; \Psi_{in}>\; \times \; a\; phase\; factor.\label{c6}
\end{equation}
We then, as usual, write the transition probability as $|T|^2$, so any phase factor is irrelevant. Then we put in factors for phase space counting many final states, and average over unselected initial states. For tachyons we ask whether that phase space is somewhat different from what it is for ordinary particles.

\section{Counting of States}

This question arises in two areas: particle scattering (looking at final states of free particles) and statistical mechanics (looking at a large equilibrium assembly of free particles.)

For ordinary particles ($v<c=1$)  we find two formulas,
\begin{equation}
d^3 p \;\;\;\;\; or \;\;\;\;\; d^3p/2E(p), \;\;\;\;\; E(p) = + \sqrt{p^2 + m^2} \ge m.\label{d1}
\end{equation}
The second formula comes from a Lorentz invariant formulation $d^4p\; \delta(p^\mu p_\mu  - m^2)$.
In particle scattering problems, one may use either formula, depending on how one normalizes the asymptotic states.  In stat mech it appears that only the first is used, even when studying highly relativistic systems. 

Now we consider tachyons ($v > c$). Here we already know to separate the 3-vector $\textbf{p}$ into a magnitude $p$ and a direction $\hat{p}$. One formula is this,
\begin{equation}
d^3 p = d^2 \hat{p}\; p^2 dp = d^2 \hat{p}\; p E dE, \;\;\;\;\; p^2 = E^2 + m^2, \;\;\;\;\;  E \ge 0.\label{d2}
\end{equation}
The Lorentz invariant formula is different,
\begin{equation}
\int d^4p \; \delta(E^2 - p^2 + m^2) = d^2 \hat{p} \;dE\;p^2/(2p), \;\;\;\;\; p (E) = +\sqrt{E^2 + m^2}\ge m.\label{d3}
\end{equation}
For scattering situations, we need to study and see which formula to use; but for stat mech there is an immediate problem with this second formula: it gives an infinite value for the average velocity of the particles. $<v/c> =<p/E> =  \int dE\; (p/2)(p/E)$.

I cannot resolve this dilemma within the present framework. In the following section we shall find an alternative path.

One more question. When we consider the Cosmic Neutrino Background being made up of tachyons, are they all in states with $E>0$ or are there an equal number of $E<0$ states? 

\section{Quantized Field for Tachyon Neutrinos}
Start by reviewing the results of my previous paper \cite{CS3}, with a slight change of notation. First, here is the result for an ordinary Dirac particle.

\begin{eqnarray}
\psi(x) =\int \frac{d^3k}{(2\pi)^{3/2}} \sum_\epsilon e^{i\epsilon(\textbf{k}\cdot \textbf{x} - \omega t)} \sum_h u_{\epsilon,h} (\textbf{k}) b_{\epsilon,h} (\textbf{k}), \label{e1}\\
u_{\epsilon,h} (\textbf{k}) = \frac{1}{\sqrt{2\omega(\omega + \epsilon m)}}\left(\begin{array}{c}  \epsilon \omega +m \\  \epsilon h k \end{array} \right) |\hat{k},h>, \label{e2}\\
 \;[b_{\epsilon,h}(\textbf{k}), b^\dagger_{\epsilon',h'}(\textbf{k}')]_+ = \delta^3(\textbf{k} -\textbf{k}') \delta_{\epsilon,\epsilon'}\delta_{h,h'},\label{e3}
\end{eqnarray}
where $\omega = +\sqrt{k^2 + m^2}$, $\epsilon = \pm 1$, $\hat{k} =\textbf{k}/k$, and $|\hat{k},h>$ is an eigenfinction of the 2-component Pauli spin matrix dotted into the direction of the momentum $\hat{k}$, with eigenvalue (helicity) $h = \pm 1$. With this construction we calculate the charge Q, derived from the conserved current $j^\mu =\bar{\psi}\gamma^\mu \psi$,
\begin{eqnarray}
&&Q = \int d^3 x\; j^0 = \int d^3 k\; \sum_{\epsilon,h}\; b^\dagger_{\epsilon,h}(\textbf{k})\;b_{\epsilon,h}(\textbf{k}),\label{e4}\\
&&[Q,b_{\epsilon,h}(\textbf{k})]_{-} = -b_{\epsilon,h}(\textbf{k}), \;\;\;\;\; [Q,b^\dagger_{\epsilon,h}(\textbf{k})]_- = +b^\dagger_{\epsilon,h}(\textbf{k}),\label{e5}
\end{eqnarray}
which lets us interpret $b^\dagger b$ as a number operator (per unit volume in momentum space). This is consistent with $d^3k$ as the density of states.

Then I constructed a quantized version of a Dirac field for a spin 1/2 tachyon, under the requirement that the fields should (anti)commute for time-like separations.  It goes like this.
\begin{eqnarray}
\psi(x) =\int_{-\infty}^\infty \frac{d\omega}{(2\pi)^{3/2}} e^{-i\omega t}\int k^2\;d^2\hat{k} \;e^{ik\hat{k}\cdot \textbf{x}} \sum_h v_h (\omega,\hat{k}) b_h (\omega, \hat{k}),\label{e6} \\
v_h (\omega,\hat{k}) = \frac{1}{\sqrt{2}\;k}\left(\begin{array}{c}  \omega +im \\  h k \end{array} \right) |\hat{k},h>, \label{e7}\\
 \;[b_h(\omega,\hat{k}), b^\dagger_{h'}(\omega', \hat{k}')]_+ = \delta(\omega - \omega')\frac{\delta^2(\hat{k} - \hat{k}')}{kk'} \delta_{h,h'}\label{e8}
\end{eqnarray}
where $k = + \sqrt{\omega^2 + m^2}$. With this construction we have,
\begin{equation}
[\psi(x),\psi^\dagger(x')]_+ = 0\;\; \;if\; (t-t')^2 > |\textbf{x} - \textbf{x}'|^2,\label{e9}
\end{equation}
which is the proper statement of causality for tachyons: No tachyon signal can propagate slower than the speed of light.

Here we calculate the flow of the conserved current $j^\mu = \bar{\psi}\gamma_5\gamma^\mu \psi$, through a surface oriented normal to the vector $\eta$,
\begin{eqnarray}
Q_\eta = \int dt\;d^2x_\perp \eta\cdot j = -\int_{-\infty}^\infty d\omega \int k^2 d^2\hat{k} \eta\cdot \hat{k} \sum_h h b^\dagger_h (\omega,\hat{k}) b_h(\omega, \hat{k}), \label{e10}\\ \;
[Q_\eta,b_{h}(\omega,\hat{k})]_{-} = +h\eta\cdot \hat{k}b_{h}(\omega,\hat{k}), \label{e11}\\ \;[Q_\eta,b^\dagger_{h}(\omega,\hat{k})]_- = -h\eta\cdot\hat{k}b^\dagger_{h}(\omega,\hat{k}).\label{e12}
\end{eqnarray}
From these equations we might interpret $b^\dagger b$ (or $bb^\dagger$) as a number operator, that is the number of particles per unit volume in momentum space; and that volume measure in momentum space is $d\omega\; k^2\;d^2\hat{k}$, which is \emph{different} from what I wrote in the previous section. 

Here is one explanation that might be offered. What I calculate above, $Q_\eta$, is an integrated flow of particles not an integrated density of particles. The difference is a factor of velocity: flow = density x velocity. So, to go from flow to density I should divide by velocity, which is $k/\omega$. That leads me to the density of states formula $d\omega\; k\;\omega\;d^2\hat{k}$. This agrees with the formula (\ref{d2}) in the previous section.

No. That is wrong. The current $j^i$ can be called a flow; but $Q_\eta$ is a time integral over that flow so it just counts the number of particles that have crossed that plane over all time. We need to ask, What is the conservation law that we rely on? For ordinary particles we have a locally conserved current $\partial_\mu j^\mu(x)= 0$; and then we prove that the 3-space integral of the time component is constant, $\frac{d}{dt} \int d^3 x\; j^0 = 0$. However, for tachyons, as discussed in \cite{CS3}, the correct integral law is different: for example, $\frac{d}{dz}\int dt\int dx\;dy \;j^3 = 0$.
That leads me to rely on the discussion above that went from the formula for $Q_\eta$ to the particle number density in momentum space:
\begin{equation}
d\omega\; k^2\; d^2 \hat{k}.
\end{equation}

An additional result from that previous study is this, 
\begin{eqnarray}
P^\nu _\eta  &=&  \int dt\; d^2 x_\perp\; \eta_\mu\; T^{\mu ,\nu} \\
& = &\int_{-\infty} ^{\infty}d\omega \int d^2 \hat{k} \;k^2\;\eta \cdot \hat{k}\sum_h h\;b^\dagger_h (\omega,\hat{k})\;b_h (\omega, \hat{k})\; (\omega, \textbf{k}).
\end{eqnarray}
This leads me to write the same momentum space density $d\omega\; k^2 \;d^2 \hat{k}$ with the additional factors $\omega$ or $\textbf{k}$ when I want to write the expressions for energy density or momentum density, respectively. That we shall do in the next section.

\section{Stat Mech for Tachyon Neutrinos}
Here is the formula [Equation (3.1.28) in reference \cite{SW}] for an ideal gas of particles of mass m, the number n(p) dp of particles of momentum between p and p+dp is given by the Fermi-Dirac and Bose-Einstein distributions at temperature T:
\begin{equation}
n(p,T,\mu) dp = \frac{4\pi g p^2 dp}{(h)^3}\left( \frac{1}{exp[(\sqrt{p^2 + m^2} - \mu)/k_B T] \pm 1}\right),\label{f1}
\end{equation} 
where $\mu$ is the chemical potential and $g$ is the number of spin states of the particle and antiparticle. For studying neutrinos in the CNB we choose the Fermi-Dirac statistics and set $\mu = 0$. We then have the following standard formulas for the particle number density $n$, energy density $\rho$ and pressure $p$:
\begin{eqnarray}
n = \frac{4\pi g}{(2\pi)^3}\int k^2\;dk/(e^{\omega/T} +1) \\
\rho = \frac{4\pi g}{(2\pi)^3}\int k^2 \omega\;dk/(e^{\omega/T} +1) \\
p = \frac{4\pi g}{3\;(2\pi)^3}\int k^2 (k^2/\omega)\;dk/(e^{\omega/T} +1),
\end{eqnarray}
where I have set the constants $c, \hbar = h/2\pi, k_B$ equal to unity for now; and I use the momentum space variables $(\omega, k)$ rather than $(E,p)$ because $p$ is commonly used to designate the pressure. For ordinary particles we have $\omega^2 = k^2 + m^2$.

For tachyons we have $k^2 = \omega^2 + m^2$; and we can always write $k\; dk = \omega\; d\omega$. Here is our first guess - TACHYON MODEL I - with the following formulas copied directly from the above.
\begin{eqnarray}
n = \frac{4\pi g}{(2\pi)^3}\int k\;\omega\;d\omega/(e^{\omega/T} +1) \\
\rho = \frac{4\pi g}{(2\pi)^3}\int k\; \omega^2\;d\omega/(e^{\omega/T} +1) \\
p = \frac{4\pi g}{3\;(2\pi)^3}\int k^3 \;d\omega/(e^{\omega/T} +1);
\end{eqnarray}
and the range of the integral over $\omega$ is $(0, \infty)$. For high energies (or $m=0$) these formulas all look the same as those above; and we have familiar results that $n \sim T^3$ and $ \rho = 3p \sim T^4$.

We are particularly interested in low energy tachyons, $\omega << k$; so we may approximate $k \approx m$. This will leave us with integrals of the form,
\begin{equation}
\int_0^\infty d\omega \; \omega^s/(e^{\omega/T} +1) = T^{s+1}\; I(s)
\end{equation}
The integrals $I(s)$ can be expressed in terms of zeta functions, but it is easier for me to calculate them directly, see Table 1.

\vskip .2cm
Table 1. \newline

\begin{tabular} {|l|l|l|} \hline
s & $\;\;\;I(s)$ &  \\ \hline 
0 & 0.693147 & ln 2 \\
1/2 & 0.678094 & \\
1 & 0.822467 & $\pi^2/12$ \\
3/2 & 1.152804 & \\
2 & 1.803085 & \\ 
5/2 & 3.082586 &\\
3 & 5.682197 & $7\pi^4/120$ \\ \hline
\end{tabular} \newline 
\vskip .2cm

For these low energy tachyons we have the total density of particles proportional to $T^2$, and the energy proportional to $T^3$, contrasting with the zero mass formulas noted above. 

The average value of the energy is $<E> = k_B T\; \frac{I(2)}{I(1)} = 2.192289\;k_BT$ 
\footnote{Using the Boltzmann distribution instead of the Fermi-Dirac gives $<E> = 2 k_B T$ for low energy tachyons, a result that  has been calculated earlier.\cite{SM}}; and the average value of the velocity ($v = m/E$ at low energies) is $<v> = (k_B T)^{-1} \frac{I(0)}{I(1)} = .842766 /k_BT$. This leads us to find $<v> = 1.847586 \frac{m}{<E>}$, which gives a correction factor to an earlier calculation of mine \cite{CS4} about Dark Energy due  to a tachyonic Cosmic Neutrino Background.

By using the results from tachyon field quantization \cite{CS3}, as presented in Section 5, we can construct an alternative set of formulas - TACHYON MODEL II - as follows.
\begin{eqnarray}
n = \frac{4\pi g}{(2\pi)^3}\int k^2 \;d\omega/(e^{\omega/T} +1) \\
\rho = \frac{4\pi g}{(2\pi)^3}\int k^2\; \omega\;d\omega/(e^{\omega/T} +1) \\
p = \frac{4\pi g}{3\;(2\pi)^3}\int k^3 \;d\omega/(e^{\omega/T} +1);
\end{eqnarray}
In the following Section we shall make physical calculations using each of these two Models.

Before leaving this section let's look at that factor $g$ that counts the number of neutrino states for each value of the momentum. By one way of counting the answer is $g=12$: there are 3 flavors of neutrinos (one each paired with the electron, mu and tau leptons) and there are two spin states and there are particle plus anti-particle. If one chooses the Weyl equation for massless neutrinos, then we get $g=6$ because each "particle" is left-handed and each "antiparticle" is right handed. If one uses the Dirac equation, for massive or massless neutrinos, then we have the full $g=12$, except that one might invoke Majorana to get back to $g=6$.  So what do I want to say for tachyon neutrinos? I have used the Dirac equation for them and found that I want to distinguish particle and antiparticle not by the sign of the frequency (Energy) but rather by the helicity. But I still have to say something about the negative frequency solutions. If I count all frequencies, then I should use $g=6$; but if I count only positive frequencies, then I should allow $g=12$. This takes us back to that question at the end of Section 4. Perhaps this matter will be settled not by theorists but by experiments.

\section{Follow the story of conventional Cosmology}
A basic set of ideas in modern cosmology  \cite{SW} is how various physical properties scale, i.e., how they vary over time with respect to the scale parameter a= a(t)  that is central to the Robertson-Walker metric of the universe. Of particular interest are the number density $n$, energy density $\rho$ and the pressure $p$ for each type of particle that is thought to be present in significant amounts

For ordinary particles and for photons their density scales as $n \sim a^{-3}$; and I will have to ask if this is true also for tachyons.

For photons, both the energy, E, and the momentum, P, scale as $a^{-1}$; so one has $\rho \sim p \sim a^{-4}$; and also $p=\rho/3$. Consistent with this is the scaling of the temperature: $T \sim a^{-1}$.

For ordinary particles of finite mass m, they behave the same as photons for $E >> m$ but then, as the universe cools, the energy E settles down to the rest energy m. So one has the energy density $\rho \sim a^{-3}$; and the pressure becomes negligible.

For tachyons of mass m, they behave the same as photons for $E >> m$. Then, as they cool, the momentum settles down to the value m while the energy continues down toward zero. To proceed further we must choose which MODEL for TACHYON stat mech to use. But first, let's outline the analysis that we will use.

At some time in the past, designated by a star notation, the tachyon neutrinos went from very high energy to very low energy. Of course this was a gradual transition but we shall imagine it as a sudden change in formulas to make our calculations easier (if not perfect). Before this transition we had tachyons behaving like photons: $T_\nu \sim T_\gamma \sim a^{-1}$, but with a constant factor relating their temperatures: $T_\nu^* = (4/11)^{1/3} T_\gamma^* = T_\gamma^*/1.401$. After the transition the photons continued to act as usual, so we can carry forward to the present time with $T_\gamma a = T_\gamma^* a^*$. Furthermore we define that transition time by writing $k_B\;T_\nu ^* = m c^2$.
Combining these formulas we have 
\begin{equation}
T_\nu^* = mc^2/k_B = T^*_\gamma/1.401 = (a/a^*) T_\gamma/1.401.
\end{equation}
We shall then use the previous formulas to see how the low energy tachyons change with temperature, relate that to the change in scale factor $a$, and get a final expressions for $n,\rho, p$ in terms of the presently measured $T_\gamma$.

TACHYON MODEL I:

We assume that, as with ordinary particles, $n \sim a^{-3}$. This says that the particle number is constant but the density decreases as space expands. The previous formula says $n \sim T_\nu^2$ so we have (dropping the constants $c,\hbar, k_B$)
\begin{equation} 
T_\nu = T^*_\nu\;(a/a^*)^{-3/2} = m \; [1.401 m/T_\gamma]^{-3/2};
\end{equation}
and the formula for pressure is 
\begin{equation}
p_\nu = \frac{4\pi g}{3\;(2\pi)^3} m^3 \;T_\nu\; I(0) \sim a^{-3/2}.
\end{equation}
Putting in numbers,  $T_\gamma = 2.725 K$ and $a=1$ for the present time, we find 
\begin{eqnarray}
MODEL \; I \\
T_\nu &=&   0.0796 \; [mc^2/0.1 eV]^{-1/2}\;\; K\\
n_\nu &=& 25.5 \;g\;\;cm^{-3} \\
p_\nu &=&    10,450\; g\;\;[mc^2/0.1 eV]^{5/2}\;\;eV\;cm^{-3}
\end{eqnarray}

TACHYON MODEL II:

Basic aspects of the formulas used here do not have the same physical meaning as those used in the first model. This is because we derived these formulas, for tachyons,  looking at the space parts of the conserved current and energy-momentum tensor, rather than using the time parts. Thus they count not the number of particle in a box but the number of particles flowing through a surface, integrated over time. This viewpoint was basic to the (non-canonical) quantization of the tachyon field theory \cite{CS3}.  This leads me to say $n \sim a^{-2}$ for low energy tachyons, which is a fundamental departure from previous thinking.  For high energy tachyons, like photons, we would expect $n \sim a^{-3}$; but at low energies we know there are other examples of behavior departing from that of photons.  The RW scale factor $a(t)$ applies to space coordinates, but not to the time coordinate; and the counting of tachyons, in Model II, follows their flow through a surface, which is 2-dimensional in space.

In this model we have $n \sim T$; so we now have $T_\nu \sim a^{-2}$. This gives us
\begin{equation}
T_\nu = T^*_\nu\;(a/a^*)^{-2} = m \; [1.401 m/T_\gamma]^{-2};
\end{equation}
and the formula for pressure is 
\begin{equation}
p_\nu = \frac{4\pi g}{3\;(2\pi)^3} m^3 \;T_\nu\; I(0) \sim a^{-2}.
\end{equation}
We also see that in this Model $\rho_\nu \sim T_\nu ^2 \sim a^{-4}$, which is the same behavior as radiation. For ordinary matter it is the number density $n_M \sim a^{-3}$ which continues, at low energy, to behave the same as radiation.

Putting in numbers, $T_\gamma = 2.725 K$ at $a=1$,  we find 
\begin{eqnarray}
MODEL \; II\\
T_\nu &=&  3.26 \times 10^{-3}  \; [mc^2/0.1 eV]^{-1}\;\;K\\
n_\nu &=& 12,840\; g\;\;[mc^2/0.1eV]\; cm^{-3} \\
p_\nu &=&    428\; g\;\;[mc^2/0.1 eV]^{2}\;\;eV\;cm^{-3}.
\end{eqnarray}

These are new results; and we can see how important they may be. This says that as the universe expands and cools, the physical property that cools the least is the pressure due to tachyons, once their energy drops below $m$.

If neutrinos had zero mass, we would have $n = 56 g cm^{-3}$, $T_\nu =1.945 K$ and $p_\nu = 0.00986 g eV/cm^3$ - very different numbers from those shown above for either Model. Low energy tachyons are remarkably different from ordinary matter and from light.

My most recent paper \cite{CS4} is about this. Looking at the existing theory of the Cosmic Neutrino Background, of a known density and temperature, one makes the guess that those neutrinos are actually tachyons with a mass of around 0.1 eV and then calculates a negative pressure throughout the universe which explains, quantitatively, the so-called Dark Energy.

Let us trace the standard theory of neutrinos in the cosmos and see where assuming they are tachyons will make a difference in the predictions. For temperatures above about $10^4 K$ (1 eV), we can take the standard model, which treats neutrinos as essentially massless Fermions (of 6 types: e, mu, tau; particle and anti-paticle). If we assume a neutrino mass (or average mass) of around 0.1 eV, then there should be some different behavior below this energy, or $10^3 K$. 

From that time to the present, photons, in the CMB, evolve as $\rho = 3p \sim a^{-4}$. Neutrinos, in the CNB, if they are ordinary particles with a mass, evolve as $\rho \sim a^{-3}$ and p is negligible. But if those neutrinos are tachyons, the story is very different: as given in either of the two Models presented above. The pressure decreases much more slowly, $p \sim a^{-1.5}$ or $p \sim a^{-2}$,  and this may become the dominant feature. This means we are facing the major task of rewriting the most recent history of cosmology. 

To do this we now follow the traditional analysis of the Friedmann Robertson Walker (FRW) model as it is used for current cosmological theory.  There is one equation, relating $\rho(a)$ and $p(a)$, which is often referred to as simply conservation of energy.
\begin{equation}
\dot{\rho} = -3 \frac{\dot{a}}{a}(\rho + p).\label{g1}
\end{equation}
Let me start by representing the pressure $p(a)$ as having contribution from several sources.
\begin{equation}
p(a) = p_R\; a^{-4} + p_T\; a^{-q} + p_\Lambda,\label{g2}
\end{equation}
where the first term comes from radiation (including any high energy particles), the second term comes from low energy tachyons (q=1.5 for Model I and q=2 for Model II) and the third term is from a postulated Cosmological Constant (CC) term in Einstein's equation: $\Lambda\;g^{\mu \nu}$ implying $\rho_\Lambda = - p_\Lambda$. 

Putting (\ref{g2}) into (\ref{g1}) leads to the solution,
\begin{equation}
\rho(a) = 3p_R \;a^{-4} + \rho_M\;a^{-3}  -\frac{3}{3-q}p_T \;a^{-q} - p_\Lambda ,\label{g4}
\end{equation}
where I have given the constant of integration a new name, $\rho_M$.
This allows a nice historical storytelling of how the universe has evolved. At much earlier times, when a was very much smaller, the first term dominated: this was the very hot period when all matter was highly relativistic. A while later, the second term came to dominate: this is the standard theory that in recent times it was baryonic matter, plus perhaps Cold Dark Matter, that comprised the majority of the energy density of the universe. Eventually, or perhaps actually now, the third term would dominate. Current theory does not consider tachyons but posits the Cosmological Constant (CC) as supplying most of the energy throughout the universe. However, if we allow consideration of tachyons, then the $p_T$ term, with $q=1.5$ or $q=2$,  could be seen as dominating in this later period, without any need to imagine a CC term. Does that minus sign in front of $p_T$ worry us?  Well, in my last paper \cite{CS4} I showed that tachyons should be expected to contribute a negative pressure as they enter Einstein's equation!

We should carry this new modeling further by looking at the other basic equation from FRW.
\begin{equation}
\frac{\dot{a}^2}{a^2} + \frac{K}{a^2}= \frac{8 \pi G }{3}\;\rho,\label{g5}
\end{equation}
where the curvature K is often set to zero. Putting (\ref{g4}) into (\ref{g5}) allows one to determine $a(t)$. If $p_T$  dominates, then we have $a(t) \sim t^{2/q}$, which becomes  an exponential on the case $q=0$.  If one uses $q=2$ for the tachyons, then this term can be said to mimic the curvature constant K.

A central success of the standard cosmology is the fitting of observational data on luminosity vs redshift for Type 1a supernovae. This uses a mixture of cold matter ($a^{-3}$) and CC ($a^0$) terms in Eq. (\ref{g4}). The substitution of the tachyon ($a^{-q}$) term for the CC term provides a plausible alternative for fitting that famous data. I am not sufficiently versed in the relevant calculations to say if this qualitative alternative is quantitatively successful, but it appears that others see some flexibility of the sort I imagine.\cite{JS}

Nevertheless, let me try and see what the numbers look like.  The critical energy density is $\rho_{crit} = 5,160 eV/cm^3$  and CC is supposed to be 0.75 of this or  $\rho_\Lambda = 3,870 eV/cm^3$

My calculation (Model II) of neutrino-tachyon pressure gives \newline $p_\nu = - 428 g [mc^2/0.1 eV]^2 eV/cm^3$ and $\rho_\nu = -3 p_\nu$. So, with g=6, I can replace CC with a neutrino tachyon mass of $m = 0.071 eV$. With Model I and g=6 it would fit $m = 0.023 eV$. 

A critique of this analysis is given in Appendix A.

\section{One More Surprise}

Lets work with the quantized field for tachyon spinors, as reported in Section 5.  We want to evaluate the conserved current and energy-momentum tensor in quantized one-particle states.

\begin{eqnarray}
j^\mu (x) = \bar{\psi}(x) \gamma_5 \gamma^\mu\psi(x), \;\;\;\;\; \label{A1}\\
T^{\mu \nu}(x) =\frac{i}{4}\bar{\psi}(x) \gamma_5 (\gamma^\mu \stackrel{\leftrightarrow}{\partial^\nu} + \gamma^\nu \stackrel{\leftrightarrow}{\partial^\mu})\psi(x).\label{A2}
\end{eqnarray}
The first thing we do is enclose these operators in $: ... :$ to denote normal ordering of annihilation and creation operators; this is done so that the vacuum expectation values of each of these operators will be zero.  This means that if I rewrite $b_1^\dagger b_2$ as $b_2 b_1^\dagger$, there will be a minus sign inserted. Following (\ref{e11}) and (\ref{e12}), I assign the operators as working for particle and anti-particle according to the helicity $h$.
\begin{equation}
b_{h=-1} (\omega, \hat{k}) |0> = 0, \;\;\;\;\;b^\dagger_{h=+1} (\omega, \hat{k}) |0> = 0;\label{A3}
\end{equation}
and I will simply write these operators as $b_{\mp} (\omega, \hat{k})$ respectively.

The construction of one-particle states is normalized as follows;
\begin{eqnarray}
\;|\omega, \hat{k},h=-1> = (k)^{1/2} b^\dagger_- (\omega, \hat{k}) |0>\label{A4} \\
\;|\omega, \hat{k},h=+1> = (k)^{1/2} b_+ (\omega, \hat{k} )|0> \label{A5}\\
\;<\omega ', \hat{k}', h' | \omega, \hat{k}, h> = \delta(\omega - \omega') \delta^2(\hat{k} - \hat{k}')
\frac{1}{k} \delta_{h h'}.\label{A6}
\end{eqnarray}

Now I calculate the expectation value of the current in any one-particle state,
\begin{equation}
<\omega, \hat{k}, h|: j^\mu(x) : \;|\omega, \hat{k}, h> = (\omega,\;\textbf{k}).\label{A7}
\end{equation}
This is nice looking because we expect this to be a 4-vector under Lorentz transformations. I believe that this verifies the choice of normalization of the one-particle states.

And for the energy-momentum tensor it comes out: 
\begin{eqnarray}
<\omega, \hat{k}, h|: T^{0,0} : \;|\omega, \hat{k}, h>  = \omega^2, \label{A8} \\
<\omega, \hat{k}, h|: T^{i,i} : \;|\omega, \hat{k}, h>  = k_i^2. \label{A9}
\end{eqnarray}

This all looks very neat, but there is one more wrinkle to consider. In my previous paper on quantizing tachyons fields \cite{CS3} I had a discussion of the technical complaint that tachyons did not allow a unitary representation of the Poincare group, except for spin zero. This has to do with looking at the "Little Group", which is O(2,1) for tachyons. There I showed that one could overcome that difficulty, for the spin 1/2 case, by introducing an indefinite metric into the Hilbert space of one-particle states: and that metric was simply the helicity operator. Doing that leads to an added factor of $h = \pm 1$ in the result for each of the inner products shown above.

For the current, that is interpreted as counting the Lepton number.  For the energy-momentum tensor it is something rather more exotic. It suggests that the neutrinos and the antineutrinos contribute as sources of gravitation with opposite signs!

In my first paper \cite{CS2} on tachyons in General Relativity I assumed the usual positive expression for $T^{\mu \nu}$ representing classical particles, This led to attractive forces among the tachyons that I imagined could lead them to accumulate into some sort of rope-like structure. If these were relatively compact and somehow attached to galaxies, I suggested, this could  produce strong localized gravitational fields that could mimic the observational effects now attributed to the mysterious Dark Matter.

In my revised paper \cite{CS4} I noted that starting from an action formulation for classical tachyons in General Relativity led to a negative sign in front of their energy-momentum tensor. This led to a picture of such tachyons, throughout the universe, being repulsed by one another but contributing a large negative pressure to the overall Robertson Walker model. This appeared to be a plausible quantitative explanation for the observational effects now attributed to Dark Energy.

We are now in the position to suggest that Cosmic Background Neutrinos (and their antiparticles) might be low energy tachyons that offer full explanations for \emph{both} of those great Dark mysteries in Cosmology.

To check the above calculations, I repeat all this for ordinary $v < c$ Dirac particles. The relevant inputs are given in the first 5 equations of Section 5. Here it is the label $\epsilon = \pm 1$, which gives the sign of the energy, that designates particle vs antiparticle. Calculating the conserved current, $j^0 = \psi^\dagger \psi$ is a positive definite quantity with any wavefunction. Applying normal ordering to the product of field operators introduces a factor $\epsilon$ to this calculation. This is what we expect: the electric charge (or lepton number or baryon number) has opposite signs for particle vs antiparticle. Then we calculate the energy-momentum tensor. This brings a differential operator $\partial^\nu$ inside the previous calculation of $j^\mu$. That adds another factor of $\epsilon$. Therefore we have $T^{00}$ and $T^{ii}$ as positive quantities, which is what we are familiar with.
Note that the metric appropriate to the Hilbert space of one-particle states for $v < c$ particles is the unit operator. For tachyons this metric was different.

There is one more bit of information to be got from the equations above. Throughout this whole paper I have been arguing with myself about what is the correct density of states expression to use for tachyons in statistical mechanics. Formula (\ref{A6}) offers a suggestion: the inverse is 
$d\omega\;d^2 \hat{k}\; k$. But this can't be correct; it does not even have the correct dimensions. This is the same as the Lorentz invariant form given in (\ref{d3}). But statistical mechanics is not concerned with Lorentz invariance. It looks at a very large collection of particles, moving in all possible directions, in a particular frame of reference where the total momentum is zero.
For ordinary particles we use $d^3k$ for the stat mech density of states; and we also use $d^3 k/\omega(k)$ for the relativistic density of states along with a normalization of one particle states as $\sqrt{\omega(k)} a^\dagger(\textbf{k})|0>$. All this leads me to reaffirm the guess that for tachyons in stat mech we should write the density of states as $d\omega\; k^2 d^2\hat{k}$, which is just what we called "Tachyon Model II" in Section 6.

\section{Experiments with Tritium}
Here we want to look at the spontaneous beta decay of Tritium (the current KATRIN experiment), and also the beta transition induced by absorption of a neutrino from the Cosmic Neutrino Background (the PTOLEMY experiment). Our objective is to see if we can make any significant predictions about these experiments on the assumption that neutrinos are tachyons rather than ordinary particles with a finite mass.

Start with the simplest task: looking at the phase space formulas for the shape of the electron energy spectrum in allowed beta decay.  Given $\Delta$ as the total energy to be given to the emitted electron plus neutrino, we have, using the $d^3 p$ formula for the neutrinos,
\begin{equation}
dN(E_e) /dE_e = const \times \int p_\nu^2 dp_\nu\; \delta(\Delta -E_e -E_\nu).\label{h1}
\end{equation}
If neutrinos are ordinary particles with a mass $m_\nu$, this gives the spectrum,
\begin{equation}
(\Delta -E_e)\sqrt{(\Delta - E_e)^2 - m_\nu^2}.\label{h2}
\end{equation}
For zero mass neutrinos this is the classic $(\Delta - E_e)^2$ formula, while for finite mass the spectrum stops at $E_e = \Delta -m_\nu$, and drops to zero with an infinite slope.

For tachyonic neutrinos, we get the spectrum,
\begin{equation}
(\Delta -E_e)\sqrt{(\Delta - E_e)^2 + m_\nu^2}.\label{h3}
\end{equation}
This goes to zero linearly at the endpoint $\Delta$.

The graph below, Figure 3,  shows these two spectra for the case of $m_\nu = 0.2 eV$ over the range of energies down to 0.5 eV below the endpoint $\Delta$. This is using Tachyon Model I, with the density of states $p^2 dp = pE dE$.

\begin{picture}(110,130)(10,20)
\put(-4,0) {\line(1,0){120}}
\put(-4,0){\line(0,1){120}}
\put(0,92){\circle*{5}}
\put(10,73){\circle*{5}}
\put(20,55){\circle*{5}}
\put(30,40){\circle*{5}}
\put(40,27){\circle*{5}}
\put(50,15){\circle*{5}}
\put(60,0.0){\circle*{5}}
\put(0,108.0){\circle{5}}
\put(10,89){\circle{5}}
\put(20,72){\circle{5}}
\put(30,56){\circle{5}}
\put(40,44){\circle{5}}
\put(50,32){\circle{5}}
\put(60,23){\circle{5}}
\put(70,15){\circle{5}}
\put(80,9){\circle{5}}
\put(90,4){\circle{5}}
\put(100,0.0){\circle{5}}
\put(95,-13){$\Delta$}
\put(60,66){\circle{5}}
\put(65,63){Tachyon $\nu$}
\put(60,83){\circle*{5}}
\put(65,80){Ordinary $\nu$}
\put(15,110){Model I}
\put(0,-20){Figure 3. Beta decay spectrum with Tachyon Model I.}
\put(123,-5){$E_e$}
\end{picture}

\vskip 2cm

The alternative, Tachyon Model II, uses the density of states $p^2 dE$ and this gives the spectrum $(\Delta-E_e)^2 +m^2$, which is depicted in Figure 4 with the same parameters.

\begin{picture}(110,130)(10,20)
\put(-4,0) {\line(1,0){120}}
\put(-4,0){\line(0,1){120}}
\put(0,92){\circle*{5}}
\put(10,73){\circle*{5}}
\put(20,55){\circle*{5}}
\put(30,40){\circle*{5}}
\put(40,27){\circle*{5}}
\put(50,15){\circle*{5}}
\put(60,0.0){\circle*{5}}
\put(0,116){\circle{5}}
\put(10,97){\circle{5}}
\put(20,80){\circle{5}}
\put(30,65){\circle{5}}
\put(40,52){\circle{5}}
\put(50,41){\circle{5}}
\put(60,32){\circle{5}}
\put(70,25){\circle{5}}
\put(80,20){\circle{5}}
\put(90,17){\circle{5}}
\put(100,16){\circle{5}}
\put(95,-13){$\Delta$}
\put(60,66){\circle{5}}
\put(65,63){Tachyon $\nu$}
\put(60,83){\circle*{5}}
\put(65,80){Ordinary $\nu$}
\put(15,110){Model II}
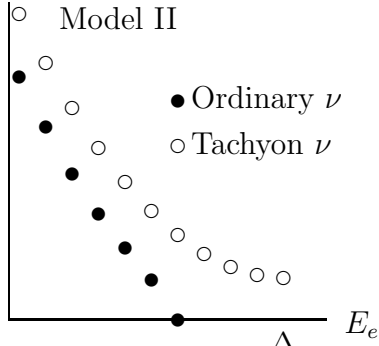
\put(0,-20){Figure 4. Beta decay spectrum with Tachyon Model II.}
\put(123,-5){$E_e$}
\end{picture}

\vskip 2cm

Avery different formalism for tachyonic neutrinos has been presented by Ciborowski and Rembielinski \cite{CR}.  They separate positive energy and negative energy solutions by means of a preferred frame of reference, which negates Lorentz invariance as usually understood. They predict an electron energy spectrum for Tritium beta-decay that looks rather like my Figure 4. Further alternative theories of neutrinos as tachyons may be found in the work of Chodos \cite{AC}.

Next we consider the transition matrix element for tachyon neutrino compared to ordinary theory.

Look at the two forms of Dirac spinors presented in Section 5: one $u_{\epsilon,h}(\textbf{k})$ for ordinary particles with a mass m and the other $v_h(\omega,\hat{k})$ for a tachyon of mass m. For large energies,  $\omega \sim k >> m$, they are the same; our interest is in low energies. For weak interactions we use the helicity projector $(1\pm \gamma_5)$ acting on these spinors and see how they compare.
\begin{eqnarray}
(1\pm \gamma_5)\;u_{\epsilon,h}(\textbf{k}) = N_o\left( \begin{array}{c} 1 \\ \pm 1\end{array} \right )\; |\hat{k},h> ,\;\;\; N_o = \frac{\epsilon \omega +m \pm \epsilon h k}{\sqrt{2\omega(\omega + \epsilon m)}},\label{h4}\\
(1\pm \gamma_5)\; v_h(\omega, \hat{k}) = N_t \left( \begin{array}{c} 1 \\ \pm 1\end{array} \right )\;|\hat{k},h> , \;\;\; N_t = \frac{(\omega + im \pm hk)}{\sqrt{2} \;k}.\label{h5}
\end{eqnarray}
So, it is just the amplitude factors that differ; and we need the absolute square of those amplitudes.
\begin{equation}
|N_o|^2 = 1 \pm hk/\omega, \;\;\;\;\; |N_t |^2= 1 \pm h \omega/k.\label{h6}
\end{equation}
What a delightfully nice result. At very low energies, like the Cosmic Background neutrinos, these factors are both close to 1. At intermediate energies, like the spontaneous beta decay of Tritium, there might be some difference, but it is not much.

We conclude that the main difference comes from the phase space and energy considerations.  For the spontaneous beta decay of Tritium the electron energy spectrum close to the end point would be one of the two graphs shown in Figure 3, or in Figure 4 above. 

For the PTOLEMY experiment, ordinary neutrinos would show a gap, amounting to $2m_\nu$, around the end point in the electron spectrum \cite{LLS}, while the tachyon neutrinos would show no such gap. What one would expect is a very narrow spike in the energy spectrum of electrons produced just at the endpoint. There is also the question of the local density of neutrinos (rather than antineutrinos) in calculating the expected rate for observation of this effect. Quite different numbers for this density are shown in Section 7 for the two Models considered there. Furthermore, if they are uniformly spread throughout the universe, one has one answer; but if they are wrapped up in bundles and attached to galaxies, then one will expect a much larger density in the Earthly location of this experiment. There has been some estimation \cite{LLS} of this effect in the case of ordinary massive neutrinos attracted by ordinary gravitational fields. However, for tachyons, within the bifurcated model presented in the last Section, I don't know how to estimate that.

\section{Summary}

In one previous paper \cite{CS3} I said that the study of tachyons needed to be extended to encompass interactions, through some generalization of S matrix theory; and in another \cite {CS4} I noted the need to fit tachyons into the theory of statistical mechanics. Both of those topics have been addressed and at least partially resolved in the present paper.

This work also leads me to make several predictions about observational consequences of the hypothesis that neutrinos, especially those that are now believed to fill the cosmos, are actually tachyons with a mass of around $0.1 eV/c^2$. The major claim is that low energy neutrinos, as tachyons, can explain Dark Energy and provide an alternative to the Cosmological Constant (CC) now inserted into Einstein's equation. My earlier paper \cite{CS4} gave a good numerical fit; and the numbers given at the end of Section 7, for either Model, remain in good agreement with experimental data. [In contrast, the vacuum energy explanation for the CC theory is off by many many orders of magnitude.] The analysis in Section 7 needs to be enlarged and incorporated into fitting of the mass of observational data\cite{P}, which now claims strong agreement with the CC model. The additional surprising minus signs found in Section 8 further suggest that Dark Matter, as well as Dark Energy might be fully explained by the gravitational forces produced by tachyonic neutrinos; however, a great deal more detailed analysis will be needed to affirm that claim. See Appendix B for some work on his topic.

The known fact of neutrino mass mixing is a complication that I have not addressed; and there remain questions about how other aspects of the standard theory of cosmology, in the more recent time span, might be revised by accepting this tachyonic hypothesis. 

I hope that, finally, I have managed to sweep aside the common prejudices against consideration of tachyons as potential physical realities (fears of negative energy states and violation of causality) and that other theorists, especially those with expertise in particle physics and cosmology, will look into this work, point out any errors, and extend the area of application and testing of these ideas.

\vskip 1 cm
\noindent{\bf Acknowledgments} 

I thank Korkut Bardakci, Jim Bogan, Robert Cahn, William Chinowski and Andrew Long for a number of helpful discussions.
\vskip 0.5cm

\vskip 0.5cm
\setcounter{equation}{0}
\def\theequation{A.\arabic{equation}}
\boldmath
\noindent{\bf Appendix A: A Lingering Problem}
\unboldmath
\vskip 0.5cm

There is something inconsistent in the work presented above.  Start with the "conservation of energy" formula,
\begin{equation}
\dot{\rho} = -3\frac{\dot{a}}{a}(\rho + p),\label{A1}
\end{equation}
involving the energy density $\rho$ and the pressure $p$ as they may depend on the scale factor $a(t)$. Now represent the pressure and energy densities as having some representation as,
\begin{equation}
\rho = \sum_q\; \rho_q/a^q, \;\;\;\;\; p = \sum_q\; p_q /a^q,
\end{equation}
we get the relations,
\begin{equation}
p_q = w_q \rho_q, \;\;\;\;\; w_q = q/3 -1;
\end{equation}
with the added note that $\rho_{q=3}$ is arbitrary.

This relation between $p$ and $\rho$ for any species is called the equation of state for that species; and here are three familiar cases: $q=4$ describes light and also any highly relativistic particles; $q=3$ describes low energy ordinary particles; and $q=0$ describes the vacuum energy (also called the Cosmological Constant). Using the standard formulas from statistical mechanics, shown at the top of Section 6 we can get those same results for the first two cases.

However, when I use the later formulas for low energy tachyons - either Model I or Model II - the relation between $p$ and $\rho$ does not fit this pattern. In Model I I found $p \sim a^{-3/2}$ and $\rho \sim a^{-9/3}$. In Model II it was $p \sim a^{-2}$ and $\rho \sim a^{-4}$.

Look again at ordinary matter as it goes from highly relativistic $\rho \sim p \sim a^{-4}$ to very slow $\rho \sim a^{-3}$ and $p \sim 0$. In that transition period can one say $p = w \rho$? In fact, looking at any of the stat mech formulas in Section 6, the relation $p = w \rho$ is possible only for these two special cases (which might be characterized as $m \rightarrow 0$ or $m \rightarrow \infty$)

If I cannot resolve this conflict, I can make this choice: take the dominant behavior, between $p$ and $\rho$ at large a (that is, choose the one with lowest value of q) and use this one to determine the other by using equation (\ref{A1}). That is in effect what I have done with the neutrino tachyon Models I and II: accept the calculated formula for p and determine $\rho$ from that via (\ref{A1}). This is consistent with the primary observation that, for low energy tachyons, the main gravitational effects come from the space-space components of the energy-momentum tensor.

\vskip 0.5cm
\setcounter{equation}{0}
\def\theequation{B.\arabic{equation}}
\boldmath
\noindent{\bf Appendix B: Model for gravitational structure of tachyons}
\unboldmath
\vskip 0.5cm

We want to follow up on the idea, first explored in \cite{CS2}, that tachyons, under mutually attractive gravitational forces, could form into some rope-like structures and then those ropes might become compact, localized entities.  The simplest model would be a circle; and the challenge is to see if there are gravitational forces capable of maintaining such a closed structure of tachyons, which all continue to move at velocities large compared to c.

In Section VI of that earlier paper I presented one simple model, which I now reject as being wrong.

Here I want to offer a more sophisticated model, which seems to offer some hope. The main point to remember is that for low energy (very fast) tachyons the primary gravitational forces come from the space components of the energy-momentum tensor. 

What I want to look at is something analogous to magnetostatics in electromagnetic theory. We know how a closed loop of wire, carrying a constant current, produces a magnetic field, and this will exert a force on any other current that is nearby. Start with the classical formula for a relativistic particle,
\begin{equation}
j^\mu (x) = \int d\tau \;\dot{\xi}^\mu (\tau)\;\delta^4 (x - \xi(\tau)) = (1, \textbf{v})\; \delta^3 (\textbf{x} - \textbf{v} t),
\end{equation}
where I have set $\xi^\mu = (\gamma \tau, \gamma \textbf{v} \tau)$ and eliminated the integral with $\delta(t - \xi ^0(\tau))$. Now I want to write an expression for the vector current density due to a uniform and time independent flow of particles along a path defined by the coordinates $\textbf{x} =\textbf{x} (s)$.
\begin{equation}
\textbf{j} (\textbf{x}) = \int ds\; \frac{d\textbf{x}(s)}{ds} \;\delta^3 (\textbf{x} - \textbf{x}(s)).
\end{equation}
One checks that this satisfies the conservation law $\nabla \cdot \textbf{j} = 0$ provided only that the path $\textbf{x}(s)$ is closed.

We now want to look at gravitation, where the source is a conserved energy-momentum tensor.
I will start by noting the simple one-particle formula for the space-space components of $T^{\mu \nu}$, which are the dominant features for low energy tachyons.
\begin{equation}
T^{i j} (x) = m \int d\tau\;\dot{\xi}^i(\tau)\dot{\xi}^j (\tau) \delta^4(x - \xi(\tau)) = 
m \gamma v_i\; v_j\; \delta^3 (\textbf{x} - \textbf{v}t).
\end{equation}
Now imagine that we have a uniform distribution of such particles moving with constant velocity along some closed path $\textbf{x} = \textbf{x}(s)$ and this is a static arrangement. This lets us ignore any retardation in calculating the field produced - just as we do in magnetostatics, with a constant current flowing in a closed loop of wire.
\begin{equation}
T^{ij} (\textbf{x}) =T_{ij} (\textbf{x}) = \int ds \; m \gamma v_i v_j \;\delta^3 (\textbf{x} - \textbf{x}(s)),
\end{equation}
where m is now understood to be the mass per unit length along the path; the total mass is $M = m \int ds$. The quantity $v_i = dx_i(s)/ds$ is the $ i^{th}$ component of the velocity of the particles along the path at the point designated by the value of $s$. (Velocity should be $dx/dt$ and s appears to represent a length rather than a time. However, our model is independent of the time and so we can choose the parameter s to be the time, in units of $c = 1$.)

Lets calculate the divergence of this tensor.
\begin{eqnarray}
\frac{\partial}{\partial x^i} T^{ij} = \int ds m \gamma \frac{dx^i(s)}{ds}\;\frac{dx^j(s)}{ds}\;\frac{\partial}{\partial x^i} \delta^3(\textbf{x} -\textbf{x}(s)) = \\
-  \int ds m \gamma \frac{dx^i(s)}{ds}\;\frac{dx^j(s)}{ds}\;\frac{\partial}{\partial x^i(s)} \delta^3(\textbf{x} -\textbf{x}(s)) = \\
- \int ds m \gamma \;\frac{dx^j(s)}{ds}\;\frac{d}{d s} \delta^3(\textbf{x} -\textbf{x}(s)) = \int ds m \frac{d \gamma v_j}{ds} \delta^3 (\textbf{x} - \textbf{x}(s)).
\end{eqnarray}
Here I have only assumed that the path closes upon itself, so that there are no end point contributions from integration-by-parts. Is this result equal to zero? We said that the magnitude of the velocity should be constant along the path; however the direction of that velocity may change. If the path is a straight line, then we do have exact conservation. But since this is a localized path that closes upon itself, there will be some measure of curvature associated with that path, call it $1/R$; and so we are left to say there is a small lack of conservation in this model so long as R is large. What do I mean by a "small" error? This problem was discussed in reference \cite{CS2} and it remains a major challenge to resolve it. Nevertheless, let us proceed.

Here is the linearized version of Einstein's equation. 
\begin{eqnarray}
g_{\mu \nu} = \eta_{\mu \nu} + h_{\mu \nu} - \frac{1}{2}\eta_{\mu \nu} h, \;\;\;\;\; h = \eta^{\mu \nu} h_{\mu \nu}, \\
\;(\frac{\partial^2}{\partial t} - \nabla ^2) h_{\mu \nu} = -16 \pi G T_{\mu \nu}.
\end{eqnarray}

We solve the (time-independent linear) equation for the metric, in the region where $v/c >> 1$
\begin{eqnarray}
g_{00} = 1 + U_0, \;\;\;\;\; U_0(\textbf{x}) = -2Gm\int ds \frac{\gamma v^2}{r}, \;\;\;\;\; r =|\textbf{r}| = |\textbf{x} - \textbf{x}(s)|,\\
g_{ij} =-\delta_{ij} +U_{ij}, \;\;\;\;\; U_{ij} (\textbf{x})= -4Gm \int ds \frac{\gamma}{r} (v_i v_j -\delta_{ij}v^2/2).
\end{eqnarray}
The expression $U_0$ looks like the potential in Newtonian gravity and it decreases as $|\textbf{x}|^{-1}$ at large distances, while $U_{ij}$ is something new. We expect that $U_{ij}$ should decrease at large distances as $|\textbf{x}|^{-3}$ \cite{CS5} ; but with the inexact vanishing of the divergence of $T_{ij}$ this will probably fail.

While I have serious questions about the calculation of $h_{ij}$ because $T_{ij}$ is not exactly conserved, $h_{00}$ should be good. And this tells us something of physical importance: If we have a localized, steady flow of tachyons, they will product an effective Newtonian gravitational field, with a total mass of $M<\gamma v^2>$. We want to see if this is a credible explanation for "Dark Matter" now ascribed to something present in the galaxies that produces the observed orbital motion of visible matter and the gravitational lensing. That expression $M<\gamma v^2>$ looks like our earlier expressions for pressure; and the magnitude of what we can expect from tachyon neutrinos for dark energy was the correct order of magnitude. The standard analysis says $\Omega_{CDM} \sim 0.25$ and $\Omega_\Lambda \sim 0.70$. So it looks quite plausible that the idea of neutrinos = tachyons can provide an alternative explanation for Dark Matter as well as for Dark Energy.

\end{document}